%% file: NUV-HD.tex
\documentclass[journal]{IEEEtran}
\input{./ds_article.def}
\input{./inst.tex}

\usepackage{tikz}
\usepackage{textcomp}
\usepackage{hyperref}
\usepackage{lipsum}

\newcommand\copyrighttext{%
  \footnotesize \textcopyright 2017 IEEE. Personal use is permitted, but republication/redistribution requires IEEE permission. See \url{http://www.ieee.org/publications_standards/publications/rights/index.html} for more information. DOI: \href{http://ieeexplore.ieee.org/document/7807295/}{10.1109/TED.2016.2641586}}
\newcommand\copyrightnotice{%
\begin{tikzpicture}[remember picture,overlay]
\node[anchor=south,yshift=25pt] at (current page.south) {\fbox{\parbox{\dimexpr\textwidth-\fboxsep-\fboxrule\relax}{\copyrighttext}}};
\end{tikzpicture}%
}

\begin{document}
\title{Cryogenic Characterization of \FBK\ \\ HD Near-UV Sensitive \SiPMs\\ }
\author{
	Fabio~Acerbi,
	Stefano~Davini,
	Alessandro~Ferri,
	Cristiano~Galbiati,
	Graham~Giovanetti,
	Alberto~Gola,
	George~Korga,
	Andrea~Mandarano,
	Marco~Marcante,
	Giovanni~Paternoster,
	Claudio~Piemonte,
	Alessandro~Razeto,
	Veronica~Regazzoni,
	Davide~Sablone,
	Claudio~Savarese,
	Gaetano~Zappal\'{a},
	and~Nicola~Zorzi
	\thanks{Manuscript revisioned on November 16, 2016.}
	\thanks{The development of the NUV-HD and NUV-HD-LF SiPM technologies was funded by the EU FP7 project SUBLIMA, Grant 241711. We acknowledge support from \NSF\ (US, Grant PHY-1314507 for Princeton University), the Istituto Nazionale di Fisica Nucleare (Italy) and Laboratori Nazionali del Gran Sasso (Italy).}
	\thanks{S.~Davini, A.~Mandarano, and C.~Savarese are with \AQGSSI\ and \AQLNGS.}
	\thanks{F.~Acerbi, A.~Ferri, A.~Gola, C.~Piemonte and N.~Zorzi are with \TNFBK\ and \TNTIFPA.}
	\thanks{M.~Marcante, V.~Regazzoni, G.~Zappal\'{a} are with \TNFBK, \TNTIFPA\ and \TNUni.}
	\thanks{C.~Galbiati is with \Princeton.}
	\thanks{G.~Korga is with \Houston\ and \AQLNGS.}
	\thanks{G.~Giovanetti, A.~Razeto and D.~Sablone are with \AQLNGS\ and \Princeton.}
	\thanks{Corresponding Author: sarlabb7@lngs.infn.it}
}
\maketitle
\copyrightnotice
\begin{abstract}
We report on the characterization of near-ultraviolet high density silicon photomultiplier (\SiPM) developed at Fondazione Bruno Kessler (\FBK) at cryogenic temperature. A dedicated setup was built to measure the primary dark noise and correlated noise of the \SiPMs\ between 40 and 300~K. Moreover, an analysis program and data acquisition system were developed to allow the precise characterization of these parameters, some of which can vary up to 7 orders of magnitude between room temperature and 40~K. We demonstrate that it is possible to operate the \FBK\ near-ultraviolet high density \SiPMs\ at temperatures lower than 100~K with a dark rate below 0.01 cps/mm$^2$ and total correlated noise probability below 35\% at an over-voltage of 6~V. These results are relevant for the development of future cryogenic particle detectors using \SiPMs\ as photosensors.
\end{abstract}
\IEEEpeerreviewmaketitle
\input{Introduction.tex}
\input{Setup.tex}
\input{Readout.tex}
\input{Analysis.tex}
\input{Results.tex}

\input{Conclusions.tex}
\bibliographystyle{IEEEtran}
\bibliography{./ds_AuthorNames}
\end{document}

%% file: inst.tex
\newcommand{\AQLNGS}{INFN Laboratori Nazionali del Gran Sasso, Assergi (AQ) 67100, Italy}
\newcommand{\AQGSSI}{Gran Sasso Science Institute, L'Aquila 67100, Italy}

\newcommand{\Houston}{Department of Physics, University of Houston, Houston, TX 77204, USA}

\newcommand{\Princeton}{Physics Department, Princeton University, Princeton, NJ 08544, USA}

\newcommand{\TNFBK}{Fondazione Bruno Kessler, Povo 38123, Italy}
\newcommand{\TNTIFPA}{Trento Institute for Fundamental Physics and Applications, Povo 38123, Italy}
\newcommand{\TNUni}{Physics Department, Universit\`a degli Studi di Trento, Povo 38123, Italy}

%% file: Introduction.tex
\section{Introduction}
\label{sec:Introduction}

\IEEEPARstart{T}{he} use of silicon photomultipliers (\SiPMs) at cryogenic temperatures is a new line of development with increasing interest from a wide range of scientific fields. \SiPM\ operation at cryogenic temperatures offers an interesting option for future particle detectors, where the detection efficiency of scintillation light is of primary importance. 

Building on its strong line of \SiPM\ development~\cite{Piemonte:2006dr}, \FBK\ recently introduced a new generation of near-ultraviolet high density (\NUVHd) \SiPMs. The performance of \NUVHd\ \SiPMs\, detailed in~\cite{Piemonte:2016cj}, makes them a promising candidate for use in future, multi-ton dark-matter detectors. One such detector is \DSk, a $20$ tonne dual phase argon time projection chamber that will require about \SI{14}{\square\meter} of near-UV sensitive photodetectors operating at \LArNormalTemperature\ with very low noise. In this paper, we present measurements of the cryogenic performance of \NUVHd\ \SiPMs\, including the dark count rate (\DCR) and the correlated noise.

We characterized two variants of the \NUVHd\ \SiPMs, the \NUVHd\ standard-field (\NUVHdSf) and the \NUVHd\ low-field (\NUVHdLf), which differ by the field strength in the avalanche region~\cite{Ferri:2016ky}. All the \NUVHd\ \SiPMs\ tested were \SI[product-units=power]{4 x 4}{\square\mm} with a cell pitch of \LNGSCryoNUVHDSPADSize.

To perform these measurements, we developed a cryogenic test setup equipped with a flexible data acquisition system and an ad-hoc analysis tool, capable of characterizing \SiPMs\ and \SiPM\ assemblies in the temperature range from \LNGSCryoSetupTemperatureRange.

Section II is devoted to the cryogenic experimental setup, Sections III and IV show the data acquisition system and the analysis tool. In Section V, we detail the results obtained for the \NUVHd\ \SiPMs.

%% file: Setup.tex
\section{Cryogenic Setup}
\label{sec:Setup}

\begin{figure}[!t]
\centering
\includegraphics[width=\columnwidth]{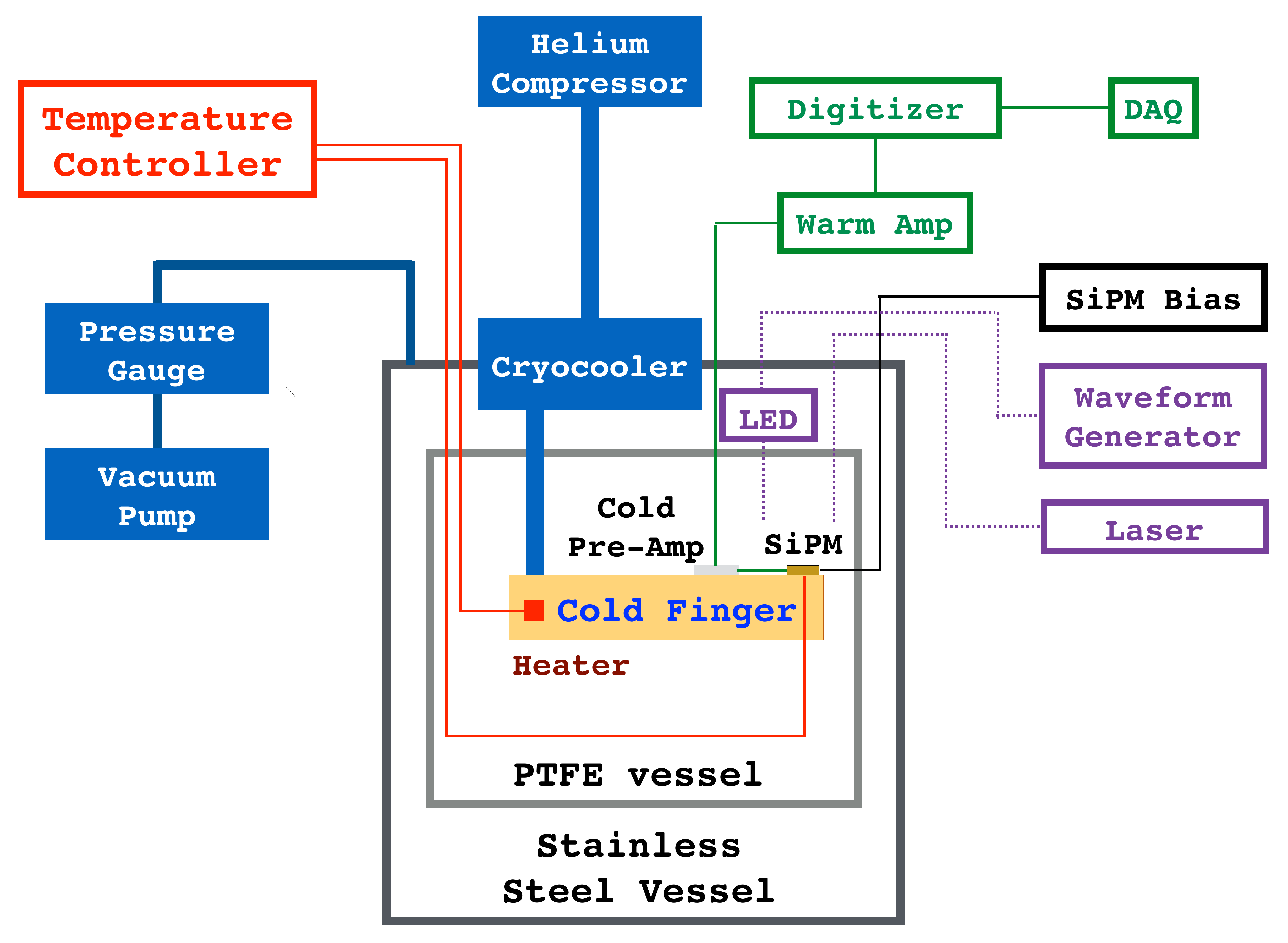}
\caption{Sketch of the experimental setup.}
\label{fig:LNGSSetup-Sketch}
\end{figure}

The cryogenic setup is contained in a stainless steel vacuum chamber made of a \LNGSCryoSetupCryostatHeight\ cylindrical tube closed by two \LNGSCryoSetupCryostatFlangeModel\ flanges.  Inside the vacuum chamber sits a thin \PTFE\ (Polytetrafluoroethylene) cylinder shielded by a layer of superinsulator that serves as a thermal shield.  The cryostat is equipped with a set of vacuum bulkhead feedthroughs. Several coaxial SMA connections are used for the SiPM signal. A set of hermetic, high density connectors are used for temperature readout, heater power, pre-amplifier power and \SiPM\ bias.  A hermetic  optical fiber feedthrough is installed so that the device under test can be illuminated by an external laser source; a blue LED is installed in the cryostat and can also be used to illuminate the \SiPM.  To minimize light leaks, which could spoil the dark rate measurements, a small dark box was built to contain the SiPM, as described later in this section.  Furthermore, opaque fabric is wrapped around the cryostat to further reduce the light entering the chamber.  An overview sketch of the system is shown in Fig.~\ref{fig:LNGSSetup-Sketch}.

During operations, the chamber reaches a vacuum level of about \LNGSCryoSetupOperatingPressure\ thanks to a \LNGSCryoSetupEvacuationPumpModel\ multi-stage roots evacuation pump.  A \LNGSCryoSetupCryocoolerModel\ pulse tube cryocooler, capable of delivering \LNGSCryoSetupCryocoolerLINPower\ of cooling capacity at \LINNormalTemperature, is mounted on the top of the flange. The cold head of the cryocooler is equipped with a cold finger, which holds the \SiPM\ assembly under test.  This arrangement allows for fast thermal cycling: the cold finger can be thermalized from room temperature to \LNGSCryoSetupTemperatureColdLimit \space in about $40$ minutes.  The cold finger also hosts a platinum surface mount RTD soldered to a printed circuit board (PCB) and connected via low thermal conductance phosphor bronze wires to a \LNGSCryoSetupTemperatureControllerModule\ temperature controller, which regulates the temperature by driving a set of high power metal film resistors mounted on the cold finger.  This system has proven to reach a temperature stability of about \SI{0.1}{\K} and an accuracy of about \LNGSCryoSetupTemperatureAccuracy\ in the range of interest from \LNGSCryoSetupTemperatureRange.  A \LNGSCryoSetupTemperatureMonitorModule\ meter measures the temperature of other platinum RTD sensors installed in the cryostat.


To minimize unwanted thermal gradients, \SiPMs\ are mounted on \LNGSCryoSetupIMSPCBThickness-thick aluminum PCBs that use insulated metallic substrate (IMS) technology with the silver-loaded conductive epoxy \LNGSCryoSetupConductiveEpoxyModel.  The mounted \SiPM\ is firmly connected to the cold finger by a screw, with a thin layer of cryogenic grease ensuring a good thermal contact between the copper and the aluminum board.  A cryogenic pre-amplifier specifically designed for this application is connected to the \SiPM\ and is in thermal equilibrium with the cold finger.


%% file: Readout.tex
\section{Readout Chain and Data Acquisition System}
\label{sec:Readout}

\begin{figure}[!t]
\centering
\includegraphics[width=\columnwidth]{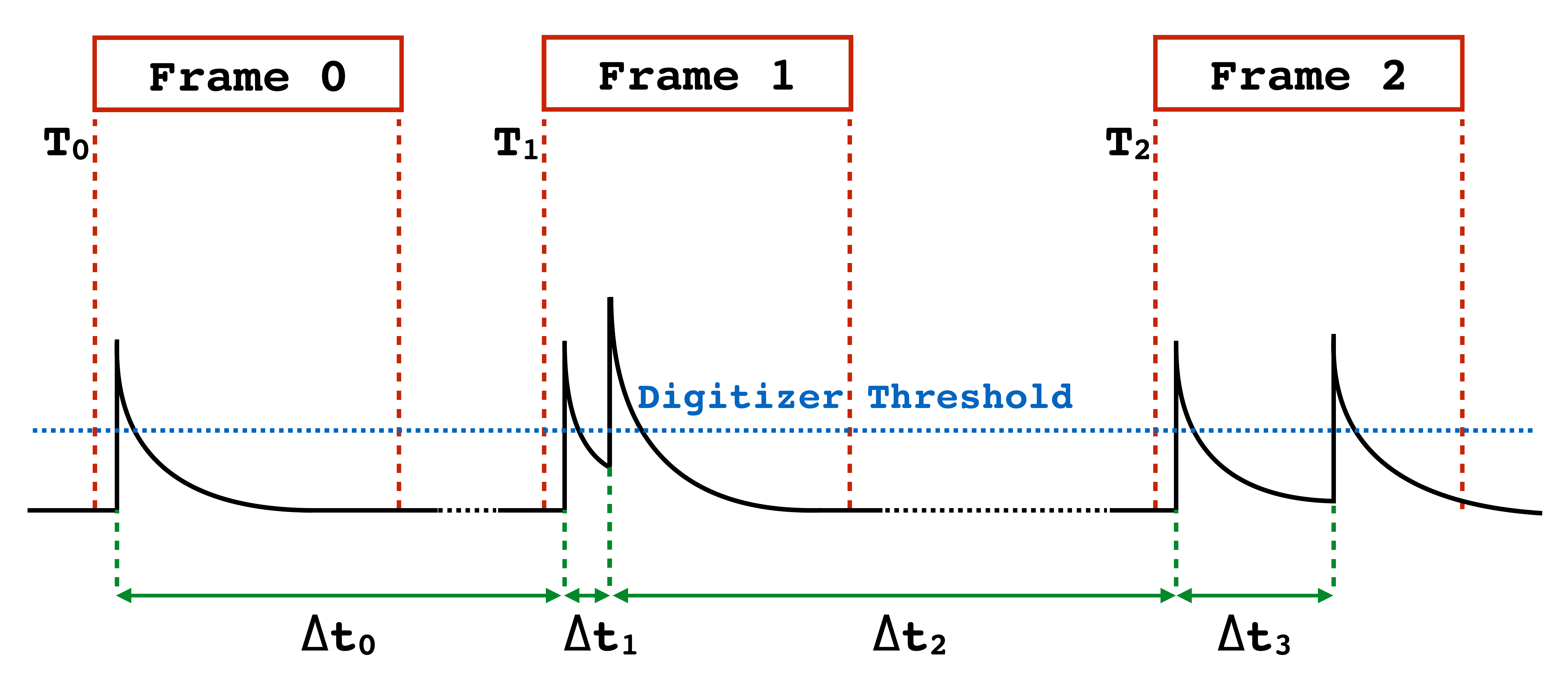}
\caption{Schematic representation of the capture of \SiPM\ waveforms in different frames.  Each frame can capture one or more \SiPM\ pulses.  See the text for details on calculation of the inter-arrival time between pulses in different frames.}
\label{fig:FrameSketch}
\end{figure}

The readout chain is composed of:
\begin{compactitem}
\item a \SiPM\ bias source;
\item a cryogenic pre-amplifier directly connected to the device under test and the bias supply;
\item a low noise warm amplifier that receives the signal from the pre-amplifier;
\item and a high speed digitizer.
\end{compactitem}

A \LNGSCryoSetupSiPMBiasModel\ is used to provide the bias voltage to the \SiPM.  This unit can directly measure the current-voltage ($I\text{-}V$) curves of the \SiPM\ useful to measure both the breakdown voltage and the value of the quenching resistor.

The cryogenic pre-amplifier is a high speed, low noise operational amplifier configured as a trans-impedance amplifier (\TIA) with a feedback resistor of \DSkTIAFeedbackResistance\, resulting in a gain of \DSkTIAGain.  This configuration allows a reasonable sensitivity despite the very large \SiPM\ input capacitance of \DSkSiPMCapacitancePerArea.  
The pre-amplifier has a \DSkTIAInputResistance\ resistor between the \SiPM\ and the \TIA\ to reduce the noise gain of the amplifier at high frequencies and to better match the signal from the \SiPM\ with the amplifier, thus avoiding undershoots~\cite{Gola:2013ke}. The design and the implementation of the cryogenic pre-amplifier will be discussed in a future publication.

The warm amplifier consists of a single stage non-inverting amplifier with input noise equivalent of \DSkWAOpAmpInputVoltageNoiseRMS, significantly smaller than the output noise from the cryogenic pre-amplifier.  The device is configured for a gain of \DSkWAGain\ with a rise time of \DSkWARiseTime.


The amplified signal is fed into a \LNGSCryoSetupDigitizerModel\ \LNGSCryoSetupDigitizerSamplingRate\ \LNGSCryoSetupDigitizerResolution\ digitizer configured for interleaved acquisition and operating in auto-trigger mode. Each time the digitizer signal crosses a fixed threshold the waveform is recorded for a fixed time duration ({\it i.e.} frame). Each saved frame is set to contain a sufficient number of points before the trigger time position to enable an estimation of the baseline level. A custom C++ program handles the configuration and the readout of the digitizer.  Data are saved in a custom file format, with each frame in the file including the digitized waveform and a header that contains accurate timing information -- see Fig.~\ref{fig:FrameSketch}. \\
Two time stamps, a fine and a coarse counter, are saved for each frame. The first is obtained directly by the digitizer and represents the starting time of each frame counting from the beginning of the measurement.  It has a resolution of \LNGSCryoSetupTimeIndicatorFineResolution\ and an overflow time of roughly \LNGSCryoSetupTimeIndicatorFineOverflow. In order to recover the time information when the acquisition time exceeds the overflow, which can easily be the case with the low event rate expected at cryogenic temperature, an additional coarse counter is extracted by the operating system of the computer with a lower precision (on the order of \LNGSCryoSetupTimeIndicatorCoarsePrecision). \\
The trigger threshold value is set by the user depending on the device under test, the noise level and the signal gain. It is typically set to be half of the value of the amplitude of a single cell response.  The length of the acquisition gate depends on the pulse rate from the \SiPM.  At cryogenic temperatures, the low rate enables the use of a short, \LNGSCryoSetupGateShortWidth\ gate, with each frame typically recording a single \SiPM\ pulse. Near room temperature, the gate is set to \LNGSCryoSetupGateLongWidth, resulting in many pulses being recorded in the same frame.  In this case, time differences between consecutive pulses are evaluated by analyzing the waveform only.

%% file: Analysis.tex
\section{The Data Analysis Software}
\label{sec:Analysis}

\begin{figure}[!t]
\centering
\includegraphics[width=\columnwidth]{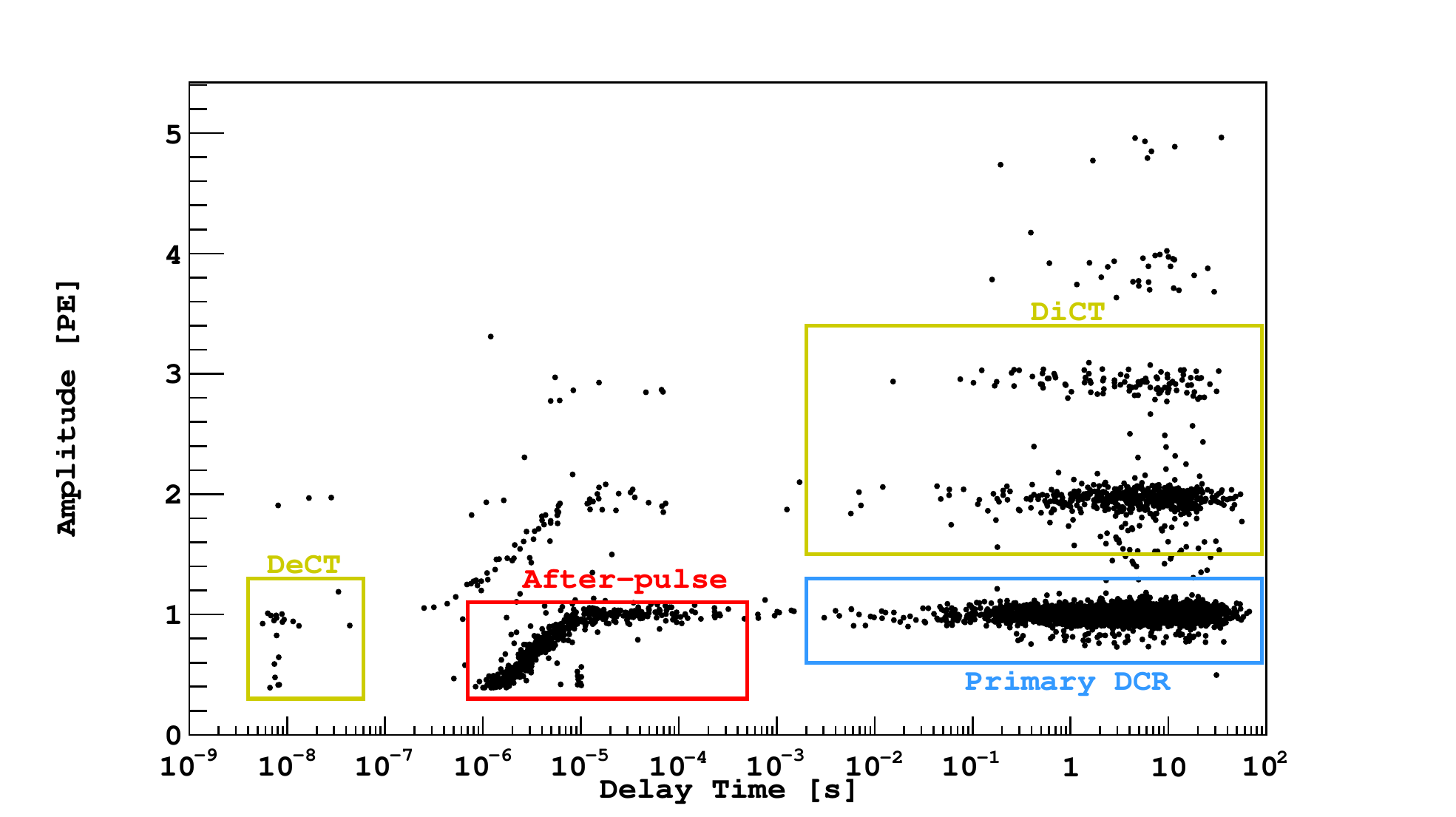}
\caption[\SiPM\ delay time versus amplitude at \LINNormalTemperature.]{Distribution of delay time versus amplitude for run at \LINNormalTemperature\ and in absence of light stimulation.  A few components of the noise response of the \SiPM\ can be clearly identified: primary dark count (part of the Dark Count Rate, \DCR), Direct CrossTalk (\DiCT), Delayed CrossTalk (\DeCT), and AfterPulsing (\AP).}
\label{fig:NUVHdLf-TCNP}
\end{figure}

\begin{figure}[!t]
\centering
\vspace{+2.5mm}
\includegraphics[width=\columnwidth]{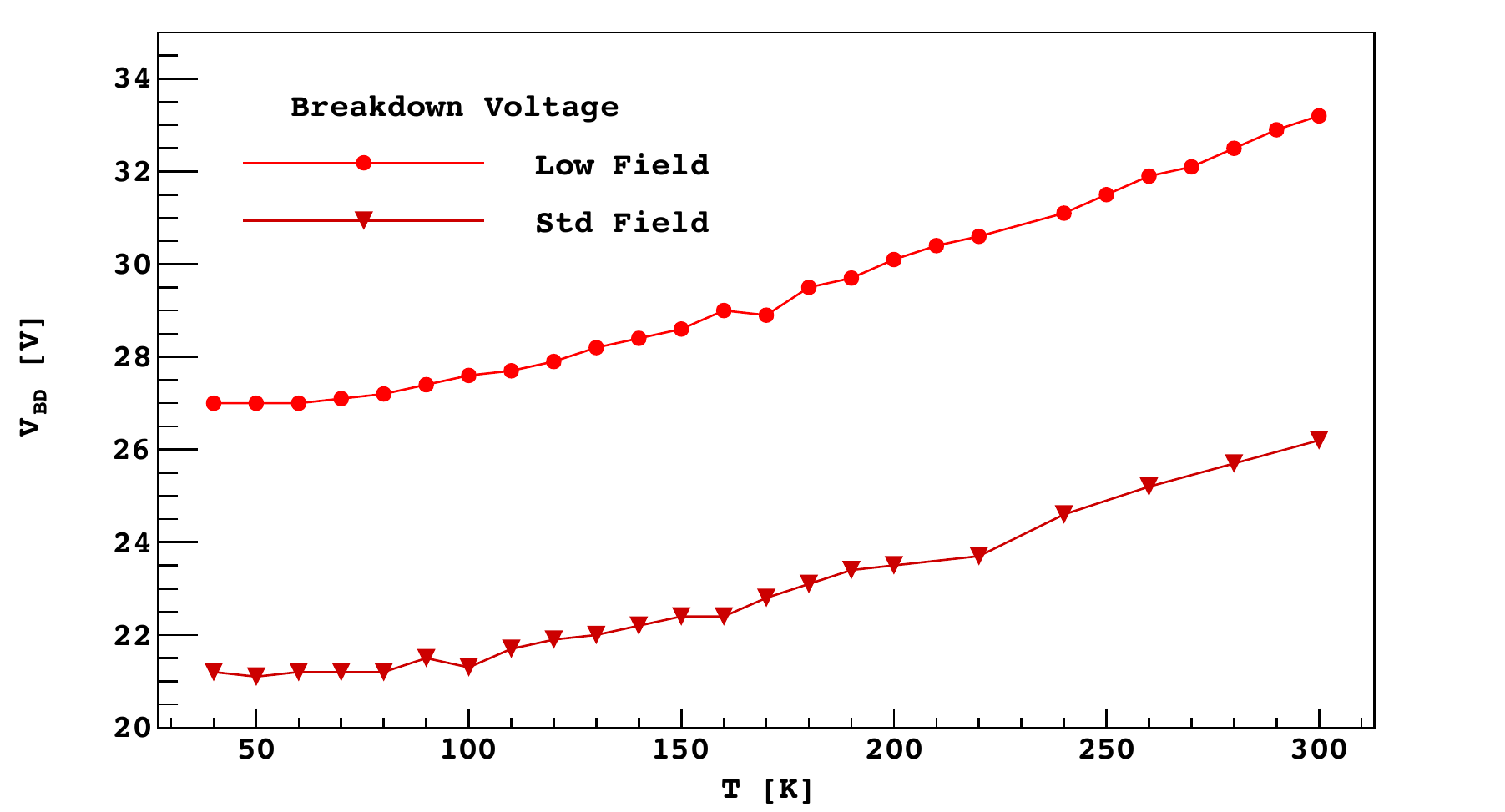}
\caption{Breakdown voltage of \NUVHdLf\ and \NUVHdSf\ \SiPMs\ as measured with a linear fit to \DLED\ single cell response amplitudes.}
\label{fig:NUVHD-Vbd}
\end{figure}

 The data analysis software was developed at \FBK\ following the procedure described in~\cite{Piemonte:2012gl}.  It reads the data stored in the data frames and performs a detailed analysis of the \SiPMs\ response, with special emphasis on the time correlation of pulses. The analysis capabilities are strongly enhanced by the addition of the time tagging of frames previously described in Sec.~\ref{sec:Readout}, which is mandated by the several order of magnitude variation in \SiPM\ dark pulse rates as a function of temperature.  The time tagging of frames allows the time difference between consecutive pulses to be calculated from the nanosecond scale, using pulses within the same frame, to thousands of seconds, using different frames, with the only practical limit set by the time available for measurement.  This allows a seamless determination of the temperature-dependent \SiPM\ pulse rate, a detailed study of their correlation, and the extraction of the secondary noise probabilities.

For data collected near room temperature, the frame time tag is not helpful, as each frame may already contain several hundreds of events and the dead time required by the digitizer to store the data becomes comparable with the inverse of the dark rate.  Except for this difference in the utilization of data, the analysis is uniform for all  dataset: each event is tagged with an ordered pair of values, {\it i.e.} the time distance from the previous pulse and the amplitude.  A typical scatter plot with $5000$ ordered pairs recorded for a run at \LINNormalTemperature\ and in absence of light stimulation is shown in Fig.~\ref{fig:NUVHdLf-TCNP}.  The plot allows the identification of the different kinds of pulses that compose the noise response of the \SiPM.

\begin{figure*}[!t]
\vspace{-5mm}
\includegraphics[width=\columnwidth]{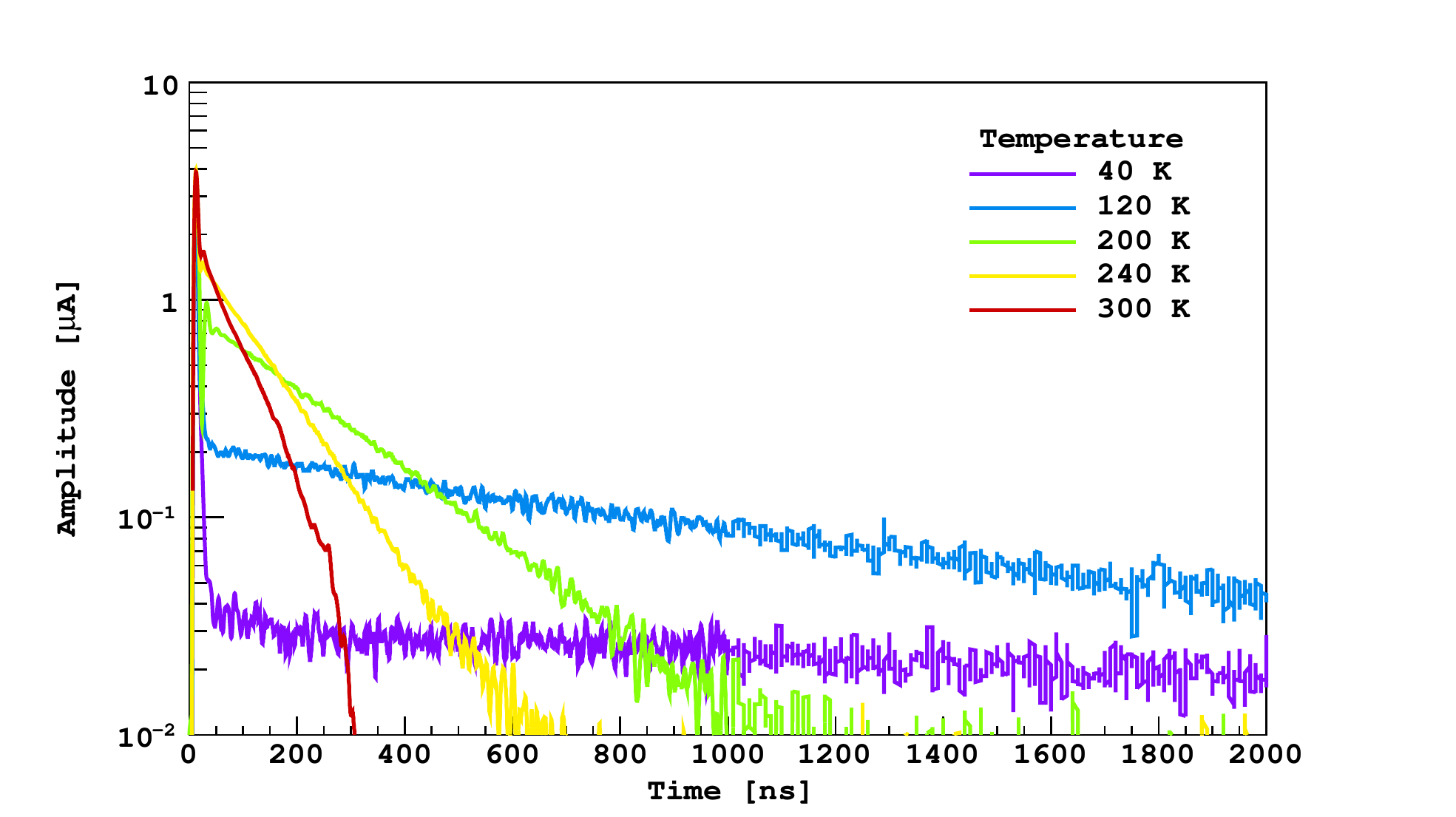}
\includegraphics[width=\columnwidth]{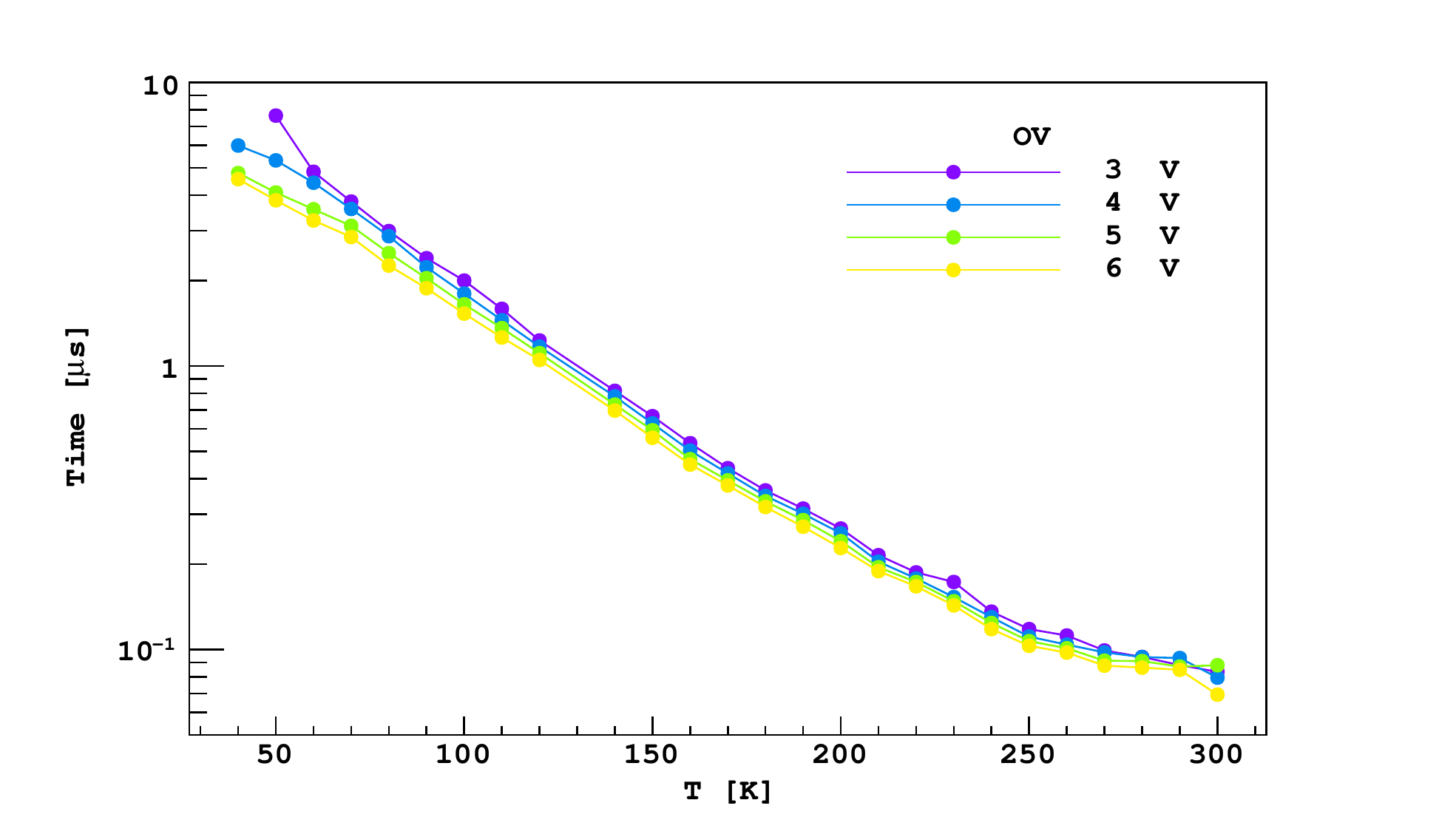}
\caption{\textbf{Left}: average of \NUVHdLf\ \SCR\ at different temperatures and at the same over-voltage.  The presence of two components at all temperatures is evident, as well as the increase of the \SPAD\ recharge time with decreasing temperatures.  A similar behavior is observed with \NUVHdSf\ \SiPMs. \textbf{Right}: \SPAD\ recharge time constant for \NUVHdLf\ \SiPMs\ as a function of over-voltage and temperature. The \NUVHdSf\ devices present the same trend but with slightly higher values.}
\label{fig:NUVHd-SCRTau}
\vspace{-5mm}
\end{figure*}

\begin{figure}[!t]
\vspace{-2mm}
\includegraphics[width=\columnwidth]{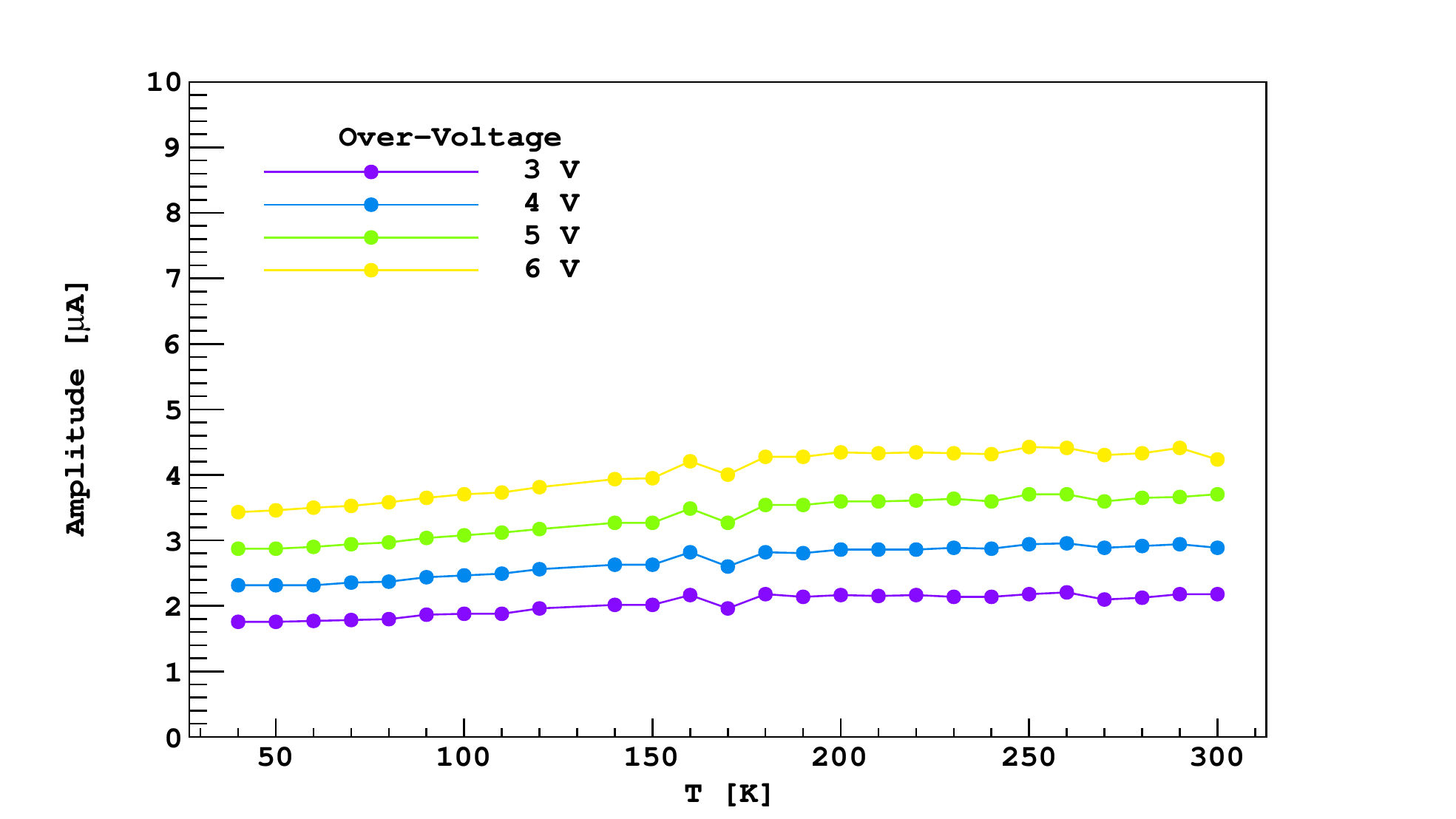}
\caption[]{Amplitude of the average \SPAD\ response as function of over-voltage and temperature for \NUVHdLf\ \SiPMs. A similar trend is found for \NUVHdSf\ \SiPMs.}
\label{fig:NUVHdLf-Ampl}
\end{figure}

\begin{asparaenum}
\item[\bf\DCR:] The main group of events is due to primary, Poisson-distributed, dark counts that make up the \DCR.  The amplitude is centered at around \SI{1}{\pe} (Photo-Electron), and the distribution in time is exponential with a decay time corresponding to the inverse of the \DCR.
\item[\bf\DiCT:] Direct CrossTalk (\DiCT) events occur a very short time after the preceding pulse, given by the travel time needed for the crosstalk photon to reach a neighboring cell and trigger an independent avalanche. The time required is on the picosecond time scale, nearly impossible to be resolved with the readout electronics.  As a result, \DiCT\ pulses are superimposed on the \DCR\ pulses and have a similar time distribution but are characterized by greater amplitudes corresponding to the detection of \num{2} or more \SI{}{\pe}.
\item[\bf\DeCT:] The less populated groups with characteristic delay times of few to tens of nanoseconds are due to Delayed CrossTalk (\DeCT).  These events are caused by crosstalk photons absorbed in the non-depleted region of a neighboring cell.  The carriers diffuse for a short time before reaching the high-field region and finally triggering an avalanche.  The resulting pulse has the single cell amplitude but is delayed with respect to the previous by the characteristic diffusion time, typically on the order of few to tens of nanoseconds.  Such delayed pulses can further trigger \DiCT\ pulses, yielding events with the same time distribution but higher amplitudes.
\item[\bf\AP:] Finally, the group of events with intermediate delay times and amplitude of \SI{1}{\pe} or lower are identified as AfterPulsing (\AP).  Afterpulsing occurs when, during an avalanche, an electron is trapped by some impurity in the silicon lattice and is then released after a characteristic time, generating a second avalanche.  Since the afterpulsing event and its primary avalanche occur in the same cell, the time distribution is determined by both the traps time constants and the recharge time constant of the microcell.  When the time distance is lower than the full micro-cell recharge, the resulting pulse has a reduced amplitude.  As with the other types of events, they can trigger \DiCT\ pulses, explaining the higher amplitude groups with the same time distribution.
\end{asparaenum}
That analysis software saves the following parameters: primary \DCR, correlated noise probabilities (\DiCT, \DeCT\ and \AP), and single cell signal features (amplitude, recovery time, and charge delivered in a fixed time gate).  Also stored is the waveform amplitude resulting from the application of the differential leading edge discriminator (\DLED) algorithm, as defined in~\cite{Gola:2012bj}, which is useful for evaluating the breakdown voltage since the resulting peak amplitude is linear with the applied over-voltage.  Measuring the \DLED\ amplitude as a function of over-voltage typically allows for a more precise determination of $V_{BD}$ (Fig.~\ref{fig:NUVHD-Vbd}) than what can be derived from the $I\text{-}V$ curve. This method allows a precise determination of the over-voltage as a function of temperature.  In order to compare the \SiPM\ \DCR\ as a function of temperature, it is important that each point is taken at the same over-voltage.  Therefore, any plot of \DCR\ vs. over-voltage is re-corrected in the analysis stage by performing a linear interpolation of the data to extract the correct \DCR\ at the exact value of over-voltage as determined with this procedure.  The same consideration applies for all the other \SiPM\ features.

The \SCR is defined as the signal generated at the output of the amplifier when a single microcell of the \SiPM\ fires. All the single photon avalanche diodes (\SPADs) contained in the \SiPMs\ under test are passively quenched with a polysilicon resistor, whose quenching resistance is denoted with $R_q$.  The impedance of this material depends strongly on temperature, increasing with decreasing temperature, and affects various signal features.  The typical \SiPM\ pulse is made up of two components, a fast one, whose time constant $\tau_f$ is of the order of few nanoseconds, and a slow component, whose time constant $\tau_s$ can extend to several microseconds and corresponds to the \SPAD\ recharge time.  The electric model proposed in~\cite{Corsi:2007is} is well suited to describe the \SiPM\ behaviour. The slow time constant is expected to be proportional to the quenching resistance, as in $\tau_s = R_q \cdot (C_{\rm SPAD} + C_q)$, where $C_q$ is the parasitic capacitance of the polysilicon resistor.  With decreasing temperature, the increase in $R_q$ translates to higher values of $\tau_s$.  This effect is clearly seen in Fig.~\ref{fig:NUVHd-SCRTau}.

\begin{figure}[!t]
\vspace{-2mm}
\includegraphics[width=\columnwidth]{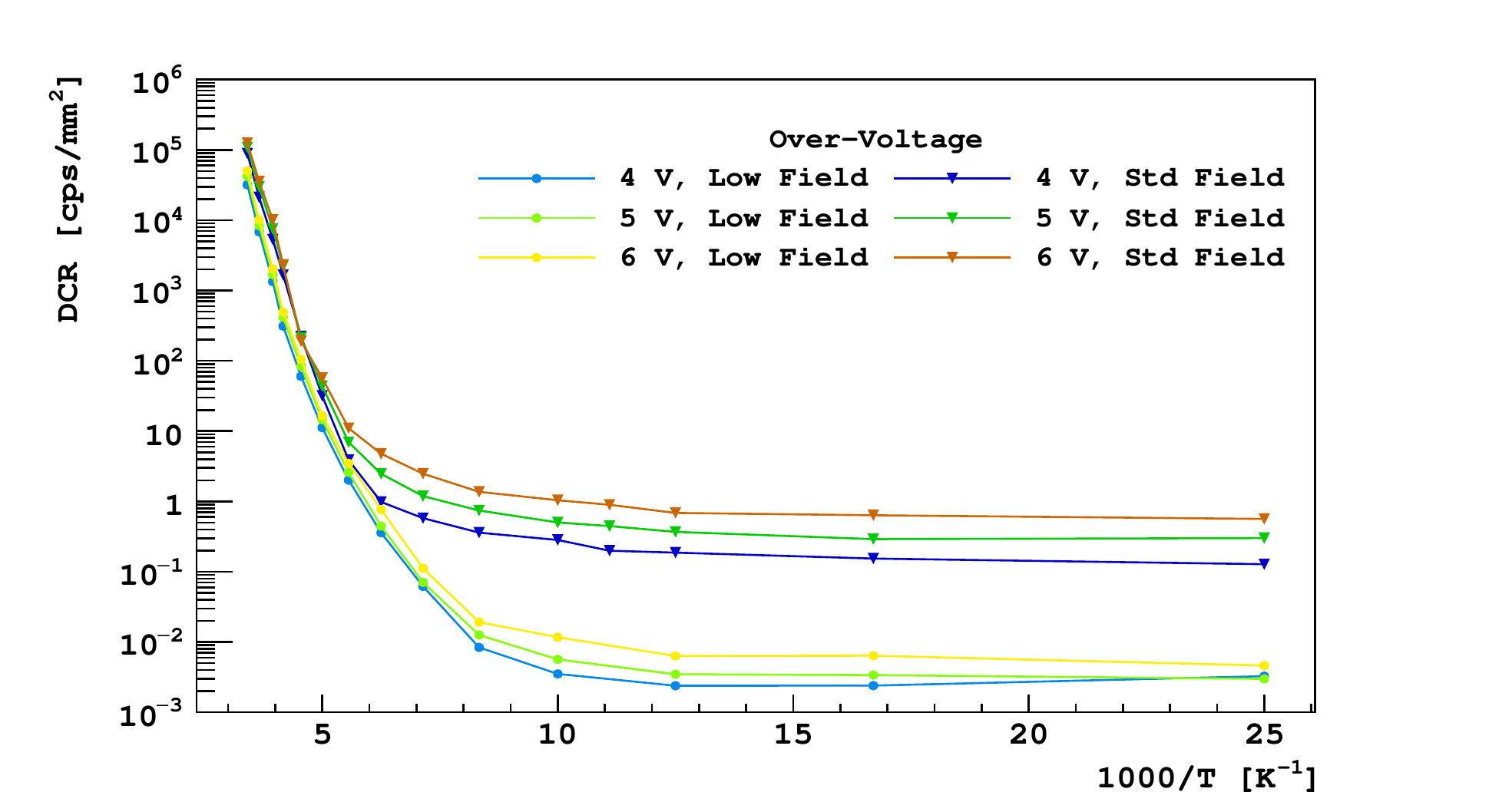}
\caption[]{\DCR\ for \NUVHdSf\ (triangular markers) and \NUVHdLf\ (circular markers) devices as function of function of 1/$T$ and over-voltage.}
\label{fig:NUVHd-DCR}
\end{figure}

%% file: Results.tex
\section{Experimental Results}
\label{sec:Results}

\begin{figure*}[!t]
\vspace{-5mm}
\includegraphics[width=\columnwidth]{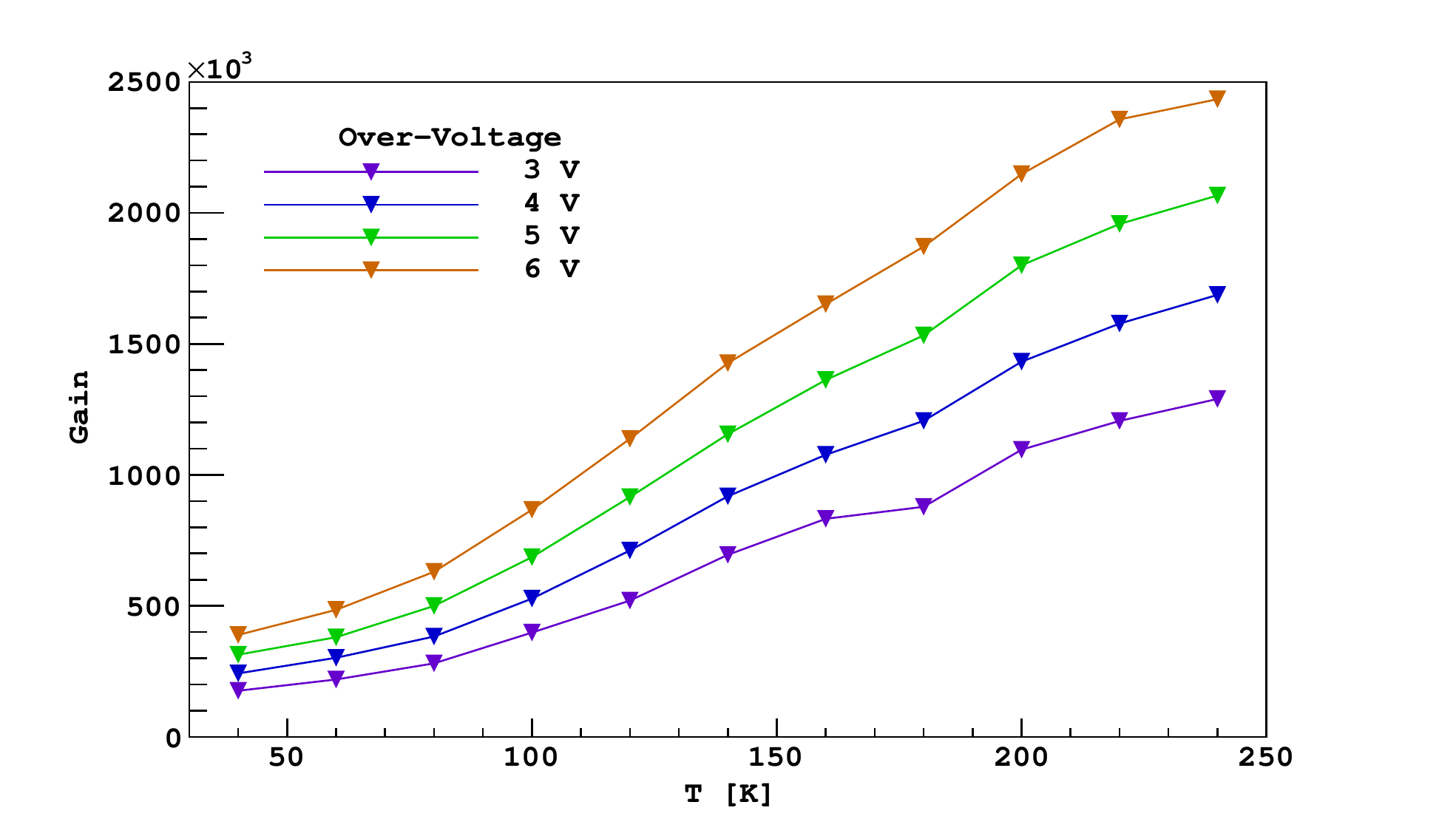}
\includegraphics[width=\columnwidth]{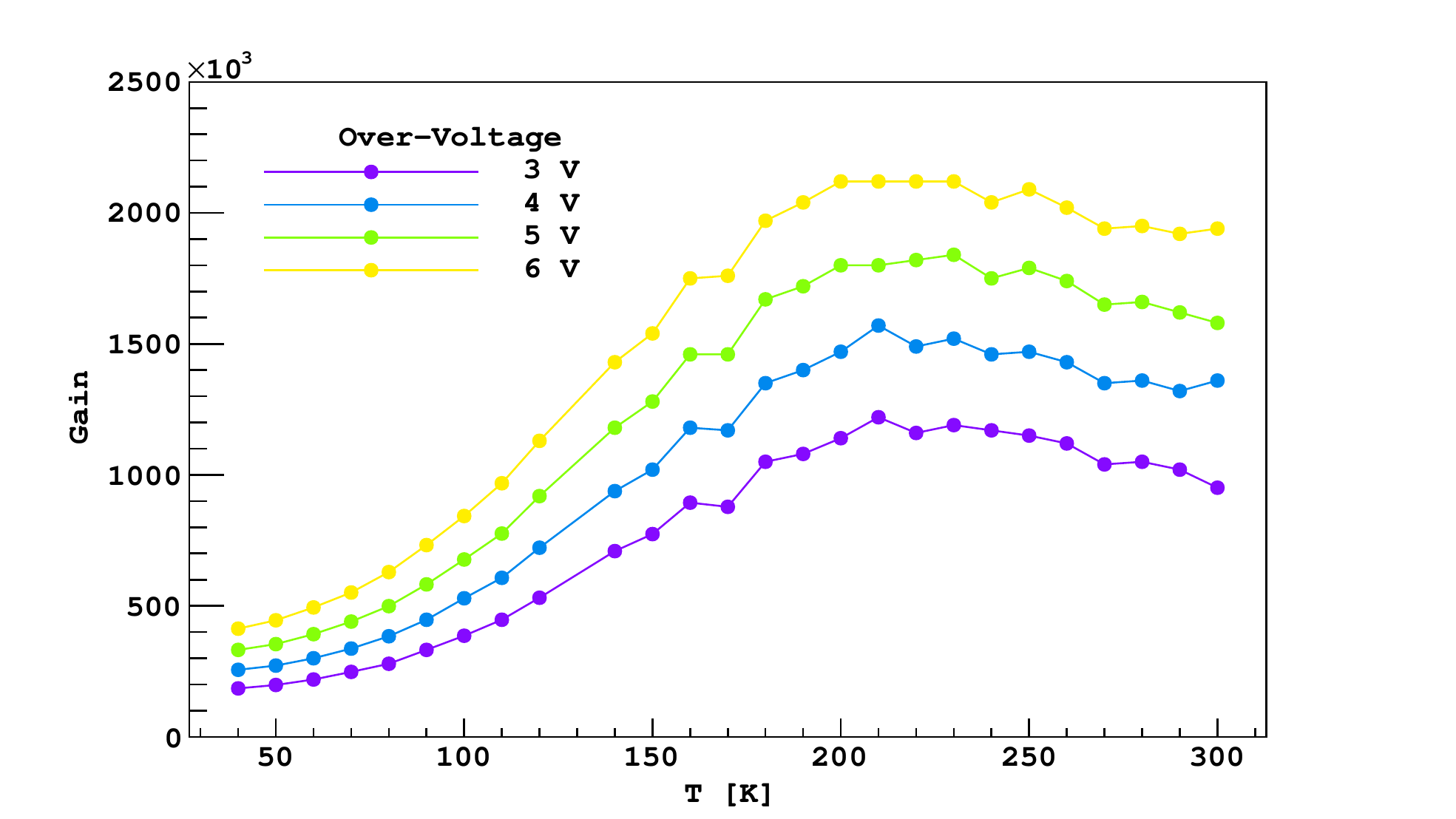}
\caption[Gain of \SiPMs\ versus over-voltage and temperature.]{Gain of the \NUVHdSf\ (left) and \NUVHdLf\ (right) \SiPMs\ in a fixed gate of \SI{500}{\nano\second} as a function of over-voltage and temperature.}
\label{fig:NUVHd-Gain}
\vspace{-5mm}
\end{figure*}

\begin{figure*}[!t]
\includegraphics[width=\columnwidth]{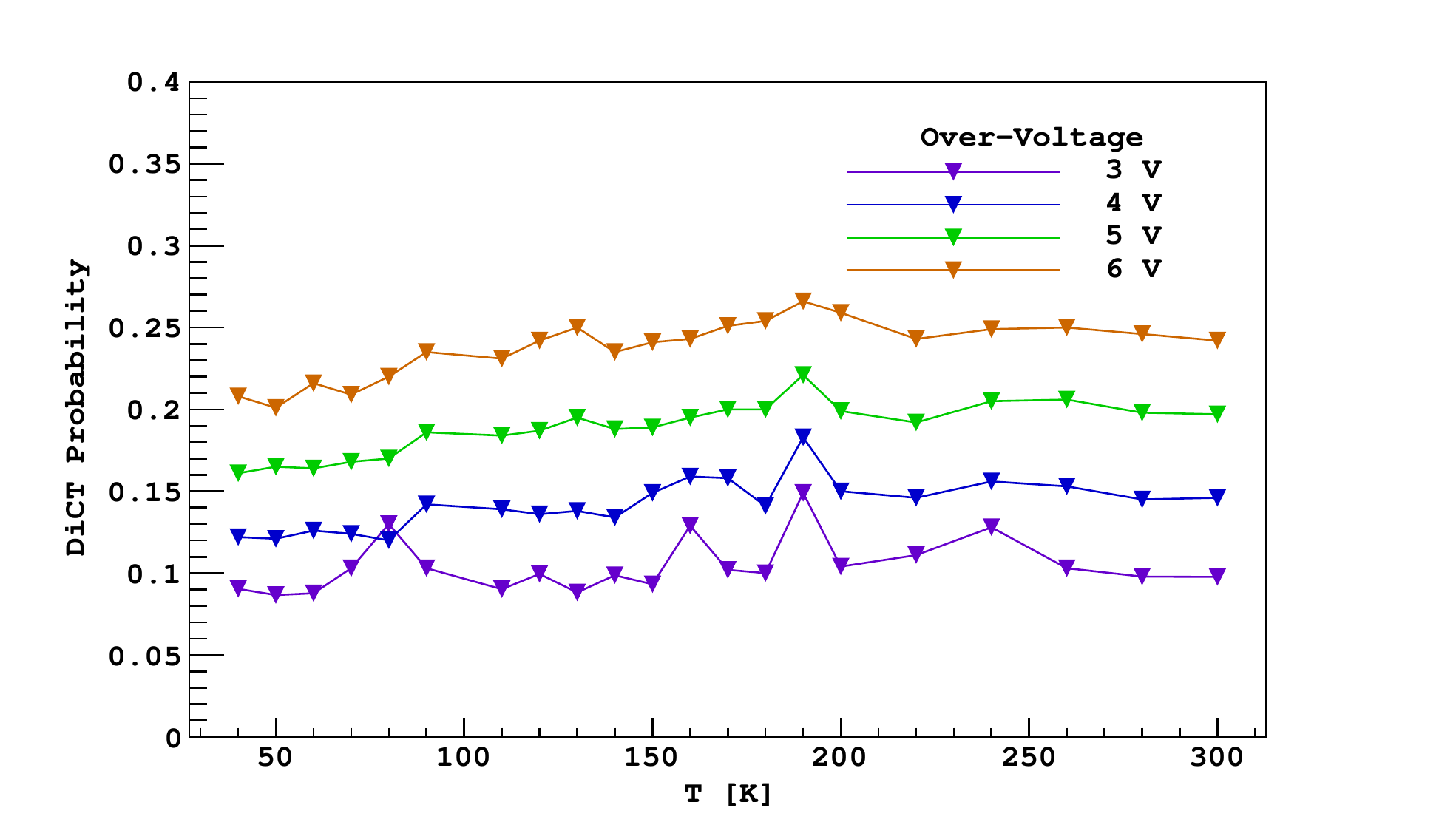}
\includegraphics[width=\columnwidth]{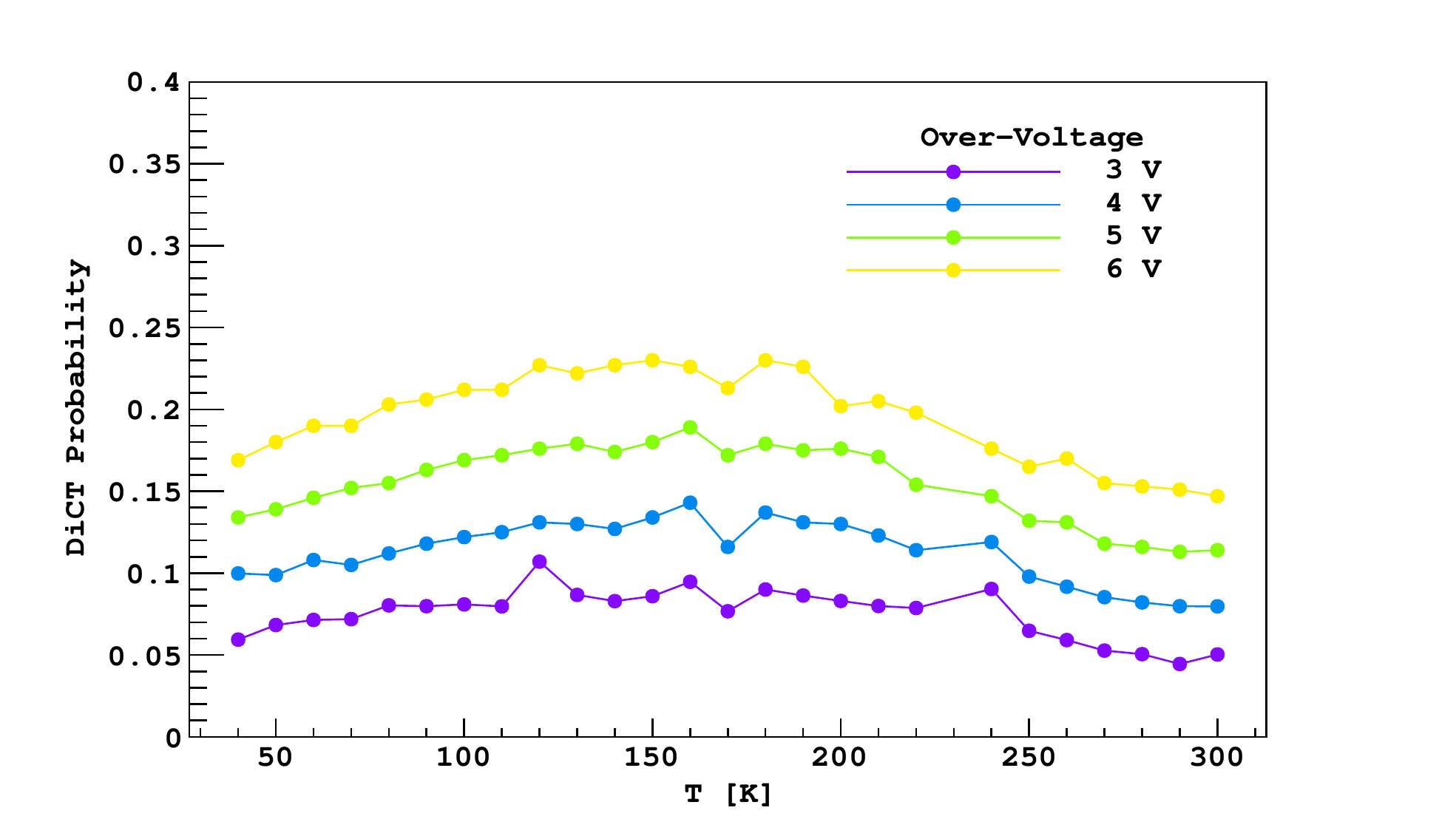}
\caption[\SiPM\ \DiCT\ probability versus temperature for different values of over-voltage.]{\DiCT\ probability as function of temperature for different over-voltage for for \NUVHdSf\ (left panel) and \NUVHdLf\ (right panel) \SiPMs.  \DiCT\ values are similar for the two technologies and exhibit a weak dependence on temperature.}
\label{fig:NUVHd-DiCT}
\vspace{-5mm}
\end{figure*}

\begin{figure*}[!t]
\includegraphics[width=\columnwidth]{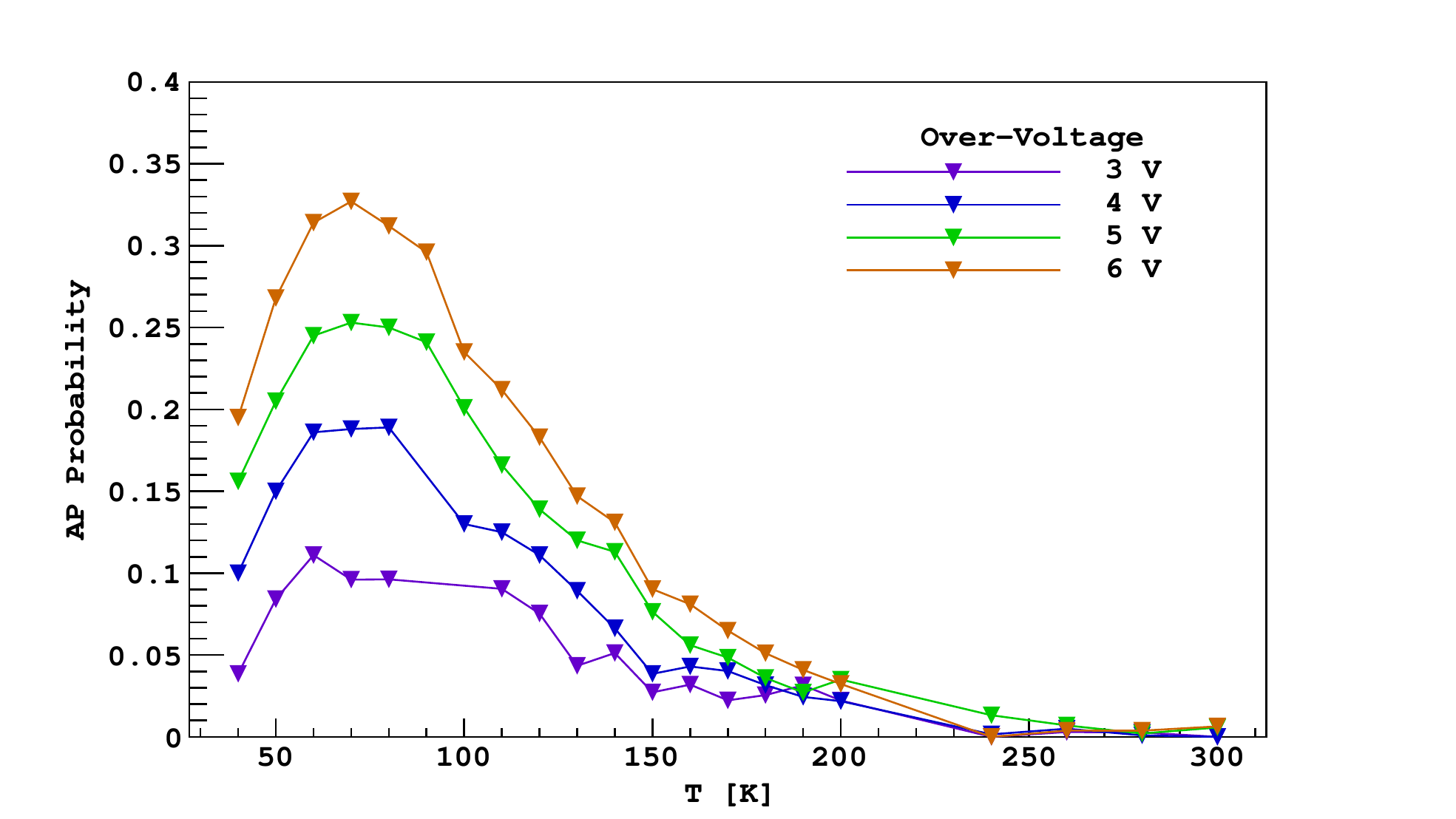}
\includegraphics[width=\columnwidth]{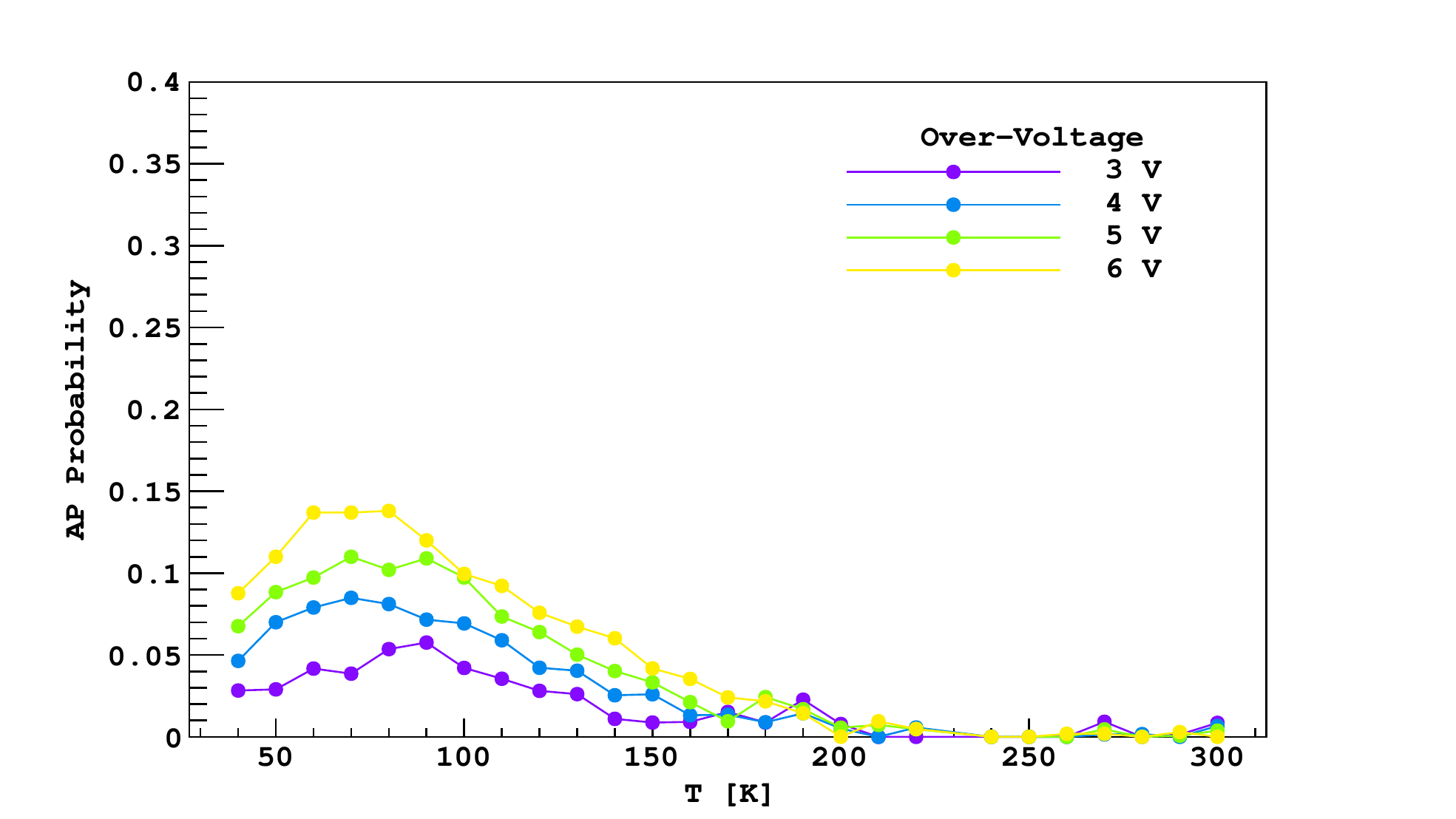}
\caption[\SiPM\ \AP\ probability versus temperature for different values of over-voltage.]{\AP\ probability as function of temperature for different over-voltage for \NUVHdSf\ (left panel) and \NUVHdLf\ (right panel) \SiPMs.  \AP\ is significantly lower for \NUVHdLf\ \SiPMs.}
\label{fig:NUVHd-AP}
\vspace{-5mm}
\end{figure*}

The analysis program described in Sec.~\ref{sec:Analysis} fits the recharge (recovery) time for the individual \SPAD\ waveform as a function of over-voltage and temperature. The result is shown in Fig.~\ref{fig:NUVHd-SCRTau}: \NUVHdSf\ and \NUVHdLf\ \SiPMs\ exhibit the same exponential trend, with a slight difference in overall results.

The peak amplitude of the \SCR\ is mainly determined by the fast peak. Its amplitude increases linearly with over-voltage and only very slowly with temperature (see Fig.~\ref{fig:NUVHdLf-Ampl}).

For each pulse, the overall charge released (or gain) is measured by comparing the result of the integration of the signal in two fixed and equal gates, one containing the pulse and one the baseline.  By measuring the distance between the peaks of the resulting distributions and considering the amplification factors of the front-end electronics, it is possible to calculate the number of  carriers generated in the \SPAD.  Again, the two variants of the \NUVHd\ technology show similar results, see Fig.~\ref{fig:NUVHd-Gain}.  The variation with temperature is due to the fact that with decreasing temperature and longer pulses, an increasingly larger fraction of the total pulse sits out of the \LNGSCryoSetupDigitizerIntegrationGate\ integration gate. Near room temperature, when the signal duration becomes shorter than the integration time, the gain is expected to saturate to a plateau. The residual dependence on temperature observed is due to the variation with temperature of the depletion width of each \SPAD\ for a given over-voltage. This happens because the breakdown voltage changes with temperature as well. Therefore, we also observe a variation of the \SPAD\ junction capacitance and thus of its gain.

The \NUVHdLf\ \SiPMs\ are optimized to achieve a significant noise reduction at low temperature and exhibit a \DCR\ that is one to two orders of magnitude lower than the \NUVHdSf\ at the same over-voltage, see Fig.~\ref{fig:NUVHd-DCR}.  An Arrhenius plot of the \DCR\ over the entire temperature range of interest clearly seperates the two different mechanisms responsible for the dark rate generation. Thermal generation dominates at high temperatures~\cite{Hall:1952iz,Shockley:1952it}, while at low temperature the main contribution originates from tunneling~\cite{Ghioni:2008ie}. Due to the lower electric field value, the magnitude of the latter component is highly suppressed in \NUVHdLf\ \SiPMs.

The \NUVHdLf\ \SiPMs\ present lower correlated noise than the \NUVHdSf\ \SiPMs\ at the same \OV, see Fig.~\ref{fig:NUVHd-DiCT}, \ref{fig:NUVHd-AP}. This is explained with the overall lower gain and avalanche triggering probability in the \NUVHdLf\ devices. The \DiCT\ exhibits only a weak dependence on the temperature, while it is linear in over-voltage, see Fig.~\ref{fig:NUVHd-DiCT}. The difference between the two technologies \AP\ probability cannot be ascribed only to the lower gain of the devices but probably also to a suppression of some field effects contributing to this noise component. The \AP\ probability reaches a maximum in the temperature range from \SIrange{60}{80}{\kelvin} and then decreases to zero in the high temperature region. This peculiar dependence can be explained by the interplay of two different phenomena having opposite effects on the noise probability. At low temperatures, the trapping time constants increase, enhancing the probability of a carrier to be released when the cell is at least partially recharged, producing an \AP\ event. On the other hand, the quenching resistance grows exponentially (see Fig.~\ref{fig:NUVHd-SCRTau}), thus suppressing the avalanche triggering probability during the recharge. The \DeCT\ probability is very low in both technologies, having a maximum value of $0.01$, and it is not shown.

%% file: Conclusions.tex
\section{Conclusions}

We developed a cryogenic setup equipped with a robust data acquisition system that is suitable for a broad variety of tests on \SiPMs\ as a function of temperature in the interval from \LNGSCryoSetupTemperatureRange.  This system is complemented by a comprehensive analysis software based on a set of tools used at \FBK\ and optimized to work in a wide range of count rates.  The \NUVHdLf\ \SiPMs\ performance at cryogenic temperature is impressive, with a \DCR\ as low as 
\SI{0.01}{cps/\square\mm}. 
Further design work for the optimization of the value of the quenching resistor as well as for the further reduction of the field in the avalanche region (and consequently of the \DCR) is ongoing at \FBK. Measurements of the PDE of the \NUVHd\ are ongoing as well as a full cryogenic characterization of the \FBK\ \RGBHd\ family of \SiPMs .